\documentclass[12pt]{article}
\usepackage{hyperref}
\usepackage{amssymb}
\usepackage{amsbsy}
\usepackage{amscd}
\ifx\boldsymbol\undefined\let\boldsymbol=\pmb\fi
\catcode`\@=11

\@addtoreset{equation}{section}
\catcode`\@=12

\def\Tr{\mathop{\rm Tr}\nolimits}
\def\diag{\mathop{\rm diag}\nolimits}
\def\btt{{\bf\tilde t}}

\newtheorem{theorem}{Theorem}
\newtheorem{remark}{Remark}
\newtheorem{Lemma}{Lemma}

\newtheorem{Corollary}{Corollary}

\def\bremark{\begin{remark}}
\def\eremark{\end{remark}}
\def\btheorem{\begin{theorem}}
\def\etheorem{\end{theorem}}

\newcount\YDcount\YDcount=0
\def\YDsize{10pt}

\def\YD#1{%
\ifnum#1=0
 \ifnum\YDcount=0 \ifx\varnothing\undefined\emptyset\else\varnothing\fi
 \else\vskip1.4pt\egroup\YDcount=0\fi
\else
 \ifnum\YDcount=0 \YDcount=1\vcenter\bgroup\vskip1pt
 \else\nointerlineskip\fi
 \vbox{\hrule\hbox{\vrule height\YDsize
 \loop\hskip\YDsize\vrule\ifnum\YDcount<#1\advance\YDcount1\repeat}\hrule
 \kern-0.4pt}\expandafter\YD
\fi}

\begin{document}
\author{A. Yu. Orlov\thanks{ Oceanology Institute, Nahimovskii
Prospect 36, Moscow, Russia, email: orlovs@wave.sio.rssi.ru} \and
T. Shiota\thanks{Mathematics Department, Faculty of Science,
Kyoto University, email: shiota@math.kyoto-u.ac.jp}}
\title{Schur function expansion for normal matrix model and associated
discrete matrix models}
\date{}
\maketitle

\begin{abstract}
We consider Schur function expansion for the partition function of
the model of normal matrices. We show that this expansion
coincides with Takasaki expansion \cite{Tinit} for tau functions
of Toda lattice hierarchy. We show that the partition function of
the model of normal matrices is, at the same time, a partition
function of certain discrete models, which can be solved by the
method of orthogonal polynomials. We obtain discrete versions of
various known matrix models: models of non-negative matrices,
unitary matrices, normal matrices.
\end{abstract}

\section{Introduction}

{\bf The model of normal matrices} (MNM) was introduced and
studied in \cite{Chau}, \cite{Zabor1} and \cite{Zabor}. A matrix
$M$ is called {\em normal}, if it commutes with hermitian
conjugated matrix:
\begin{equation}
[M,M^\dag]=0
\end{equation}
One can bring the matrix $M$ to
its diagonal form via the unitary matrix $U$: $M=UZU^\dag$, where
$Z=\diag(z_1,\dots,z_n)$, with $z_i$ the eigenvalues of $M$.
It is clear that $M^\dag=U{\bar Z}U^\dag$, where the diagonal matrix ${\bar
Z}$ is the complex conjugate of $Z$.

The model of the normal matrices is defined by its partition function as
follows:
\begin{equation}\label{normint}
Z_n=C\int d\Omega(M)
e^{\Tr V_1(M)+\Tr V_2(M^\dag)+\Tr V(M,M^\dag)} \ ,
\end{equation}
where $C$ is a normalization constant. The integration measure is
defined as
$$ 
d\Omega(M)=d_*U|\Delta(z)|^2\prod_{i=1}^n d^2z_i \ ,
$$
where $d_*U$ is the Haar measure on the unitary group $U(n)$,
and $\Delta(z)$ is the Vandermonde determinant:
$$ 
\Delta(z)=\det\bigl(z_i^{n-k}\bigr)_{i,k=1,\dots,n}=\prod_{i<k}^n(z_i-z_k)
$$
For the case $n=0$, one sets $\Delta(z)=1$.

The potentials $V_1(M)$, $V_2(M^\dag)$ are defined by their Taylor series as
follows:
$$ 
V_1(M):=\xi({\bf t},M):=\sum_{m=1}^\infty t_mM^m,\quad
V_2(M^\dag):=\xi({\bf t'},M^\dag)=\sum_{m=1}^\infty
t_m'\left(M^\dag\right)^m \ ,
$$
where ${\bf t}=(t_1,t_2,\dots)$, ${\bf t'}=(t_1',t_2',\dots)$;
$t_m$, $t_m'$ are called coupling constants. The interaction
term $V$, a function of two variables, is not specified here.
Integral (\ref{normint}) is supposed to be convergent; in the
present paper, we consider it as its perturbation series in
the coupling constants.

After the change of variables $M\to (U,Z)$ and the integration
over $U(n)$, one obtains
\begin{equation}\label{normint-}
Z_n=Z_n(V;{\bf t},{\bf t'})=\int_{\mathbb C} \cdots
\int_{\mathbb C} |\Delta(z)|^2 \prod_{i=1}^n e^{V_1(z_i)+V_2({\bar
z}_i)}e^{V(z_i,{\bar z}_i)}d^2z_i \ ,
\end{equation}
where the integration is over the complex planes of eigenvalues
$z_i$. Here we choose the normalization constant $C=C(n)$ in such
a way that $C\cdot\mathop{\rm Vol}(U(n))=1$.

The model of normal matrices has applications in a description of
quantum Hall droplets. As it was shown in \cite{MWZ}, the
particular case of the model, namely, the case when $V(z,{\bar
z})=-|z|^2$ in its $n\to \infty$ limit, is related to the
interface dynamics of water spot with a constant source surrounded
by oil; in addition, it is related to certain old problems
of complex analysis. The relation of these problems to Toda
lattice equations \cite{UT} and its dispersionless analogue
\cite{TT} was found and discussed in \cite{MWZ}.

The integrable structure of the model of normal matrices was
found and discussed in \cite{Zabor}.
The coupling constants ${\bf t}$, ${\bf t'}$ and the size of matrices
$n$ were identified with the so-called Toda lattice higher times.

\bremark\label{V1V2intoV}
One can absorb both $V_1$ and $V_2$ into the term $V$.
We prefer not to do so, since
we are interested in the integrable structure of the model; we
therefore keep the parameters ${\bf t}=(t_1,t_2,\dots)$,
${\bf t'}=(t_1',t_2',\dots)$ as the Toda lattice higher times, see
below. \eremark

\bremark If the term $V$ enjoys axial symmetry
\begin{equation}\label{axial}
V(z,{\bar z})=V_{\rm axi}(|z|^2) \ ,
\end{equation}
then, for a certain choice of coupling constants, there exists a
different integrable structure: partition function (\ref{normint})
is an infinite-soliton tau function of a ``dual'' TL hierarchy,
higher times of the dual hierarchy being related to the moments of
$e^{V_{\rm axi}}$. See \cite{hypsol} about this and related topics.
\eremark

\paragraph*{Schur functions and partitions} Polynomial functions in many
variables, like the Schur functions, are parameterized by
partitions. A {\em partition} is a sequence of non-negative integers
in the non-increasing order:
\begin{equation}\label{partition}
\lambda = (\lambda_1, \lambda_2, \dots,\lambda_r,\dots)\ ,\quad
\lambda_1 \ge \lambda_2 \ge \dots \ge\lambda_r \ge \dots,
\end{equation}
where we identify $(\lambda_1,\dots,\lambda_n)$ with
$(\lambda_1,\dots,\lambda_n,0)$, and in the case of infinite sequence,
assume $\lambda_r=0$ as $r\gg0$.
The $\lambda_i$ in (\ref{partition}) are called the {\em parts} of
the $\lambda$. The number of nonzero parts is the {\em length} of
$\lambda$, denoted by $\ell(\lambda)$. The sum of the parts is the
{\em weight} of $\lambda$, denoted by $|\lambda|$. If
$n=|\lambda|$, we say that the $\lambda$ is a {\em partition of}
$n$. The partition of zero (where $\lambda_1=0$) is denoted by
boldface $\bf 0$, to distinguish it from number 0.
The set of all partitions, including $\bf0$, is denoted by $P$.

The {\em (Young) diagram} of a partition $\lambda$
is defined as the set of points (or nodes) $(i,j) \in {\mathbb Z}^2$, such
that $1\le j \le \lambda_i$. Thus, it is a subset of a rectangular array
with $\ell(\lambda)$ rows and $\lambda_1$ columns. We denote the diagram
of $\lambda$ by the same symbol $\lambda$. For example,
\begin{equation}\label{YD331}
\YD3310
\end{equation}
is the diagram of $(3,3,1)$. The weight of this partition is $7$,
the length is equal to $3$.

The partition whose diagram is obtained by the transposition of
the diagram $\lambda$ with respect to the main diagonal is called
the conjugated partition and denoted by $\lambda^t$.

The {\em product of hook lengths} $H_\lambda$ is defined as
\begin{equation}\label{Hlambda}
H_\lambda=\prod_{i,j\in \lambda}h_{ij},\quad h_{ij}= \lambda_i-i
+\lambda_j^t-j+1 \ ,
\end{equation}
where the product ranges over all nodes of
the diagram of the partition $\lambda$.

Given number $q$, the so-called hook polynomial $H_\lambda(q)$ is
defined as:
\begin{equation}\label{Hlambda(q)1}
H_\lambda(q)=\prod_{i,j\in\lambda}(1-q^{h_{ij}}),\quad h_{ij}=
\lambda_i-i +\lambda_j^t-j+1
\end{equation}
In what follows, we also need notations:
\begin{equation}\label{n(lambda)}
n(\lambda) :=\sum_{i=1}^k (i-1)\lambda_i \ ,
\end{equation}
\begin{equation}\label{Poch}
(a)_\lambda :=
(a)_{\lambda_1}(a-1)_{\lambda_2}\cdots(a-k+1)_{\lambda_k} \
, \quad (a)_m :=\frac{\Gamma(a+m)}{\Gamma(a)} \ ,
\end{equation}
\begin{equation}\label{Pochq}
(q^a;q)_\lambda :=
(q^a;q)_{\lambda_1}(q^{a-1};q)_{\lambda_2}\cdots(q^{a-k+1};q)_{\lambda_k}
\ , \quad (q^a;q)_m :=(1-q^a)\cdots(1-q^{a+m-1}) \ ,
\end{equation}
where $k=\ell(\lambda)$. We set $(a)_0=1$ and $(q^a;q)_0=1$.

We now consider a semi-infinite set of variables
${\bf t}=(t_1,t_2,t_3,\dots)$. Given partition $\lambda$, the Schur
function $s_\lambda({\bf t})$ is defined by
\begin{equation}\label{Schurt}
s_\lambda({\bf t})=\det\bigl(h_{\lambda_i-i+j}({\bf t})\bigr)_{1\le i,j\le
\ell(\lambda)}\ ,\quad\hbox{where}\quad
\sum_{k=0}^\infty z^kh_k({\bf t}) =
\exp\sum_{m=1}^\infty z^mt_m
\ ,
\end{equation}
and, for $k<0$, we put $h_k=0$ . The $h_k({\bf t})$ is called the
elementary Schur function.

There is another definition of the Schur function; it is the
following symmetric function in the different variables
$x:=x^{(n)}:=(x_1,\dots,x_n)$, where $n\ge\ell(\lambda)$:
\begin{equation}\label{detSc}
\underline s_\lambda(x)=\frac{\det\bigl(x_i^{\lambda_j-j+n}\bigr)_{1\le
i,j\le n}} {\det\bigl(x_i^{n-j}\bigr)_{1\le i,j\le n}} \ ;
\end{equation}
for the zero partition one puts $\underline s_{\bf0}(x)=1$.
If
$$ 
{\bf t}={\bf t}(x^{(n)})=(t_1(x^{(n)}),t_2(x^{(n)}),\dots),\quad t_m(x^{(n)})=\frac 1m \sum_{i=1}^n x_i^m,
$$
then
definitions (\ref{Schurt}) and (\ref{detSc}) are equivalent \cite{Mac}:
\begin{equation}
s_\lambda({\bf t}(x^{(n)}))=\underline s_\lambda(x^{(n)}).
\end{equation}
\bremark\label{Schurvanish}
From definition (\ref{Schurt}) it follows that
$s_\lambda({\bf t}(x^{(n)}))=0$ if $\ell(\lambda)>n$. \eremark
The Schur functions $\underline s_\lambda(x_1,\dots,x_n)$, where
$\ell(\lambda)\le n$, form a basis in the space of symmetric
functions in $n$ variables.
We use the underline in $\underline s_\lambda$ only to distinguish
the two definitions. If an $n\times n$ matrix $X$ has eigenvalues
$x_1$, \dots, $x_n$, we may denote $\underline s_\lambda(x_1,\dots,x_n)$
by $s_\lambda(X)$, without underline, since in this paper the Schur
function with uppercase argument is used only in this sense.
\begin{remark}
If any of $x_i$ coincides with any of $ \bar y_i $, say, $x_k=\bar y_j$,
then we assume that
the function $e^{V(z,{\bar z})}$ vanishes at $z=x_k^{-1}$ so that
the integral (\ref{normint-}) is convergent.
\end{remark}

\paragraph*{Soliton theory} While the KP hierarchy of equations,
\cite{ZSh}, \cite{JM}, \cite{D}, is the most popular example in
the soliton theory, the two-dimensional Toda lattice (TL) is
another important equation, first integrated in \cite{AM}, and
carefully studied in \cite{UT} in the framework of \cite{JM}:
\begin{equation}\label{Toda}
\partial_{t_1}\partial_{t'_1}\varphi_n=e^{\varphi_{n+1}-\varphi_n}-
e^{\varphi_n-\varphi_{n-1}}
\end{equation}
This equation gives rise to the TL hierarchy which contains
derivatives with respect to the higher times $t_1$, $t_2,\dots$ and
$t_1'$, $t_2',\dots$\,. This equation was applied to the study of
models of random Hermitian matrices (the so-called one- and
two-matrix models) in \cite{GMMMO}, \cite{ZKMMO}.

The key point of the soliton theory is the notion of tau function,
introduced by Sato (for KP tau function; see \cite{JM}). The tau
function, a sort of a potential which gives rise to the TL hierarchy,
is a function of the two sequences of higher times,
$t_1,t_2,\dots$ and $t_1',t_2',\dots$\,, and the discrete variable $n$:
$\tau=\tau(n,{\bf t},{\bf t'})$. More
explicitly, we have \cite{JM}, \cite{UT}:
$$ 
\quad u=2\partial_{t_1}^2\ln \tau(n,{\bf t},{\bf t'})\ ,\quad
\varphi_n({\bf t},{\bf t'})=\ln \frac{\tau(n+1,{\bf t},{\bf t'})}
{\tau(n,{\bf t},{\bf t'})}
$$

\paragraph*{Power series solutions}

The formal power series solutions of the TL hierarchy were
described by Takasaki in the form of double series in the Schur functions
over partitions \cite{Tinit} (see also \cite{TI}, \cite{TII}):
\begin{equation}\label{tauschurschur}
\tau(n,{\bf t},{\bf t'})=\sum_{\lambda,{\lambda'} \in P}g_{\lambda,
{\lambda'}}(n)s_\lambda({\bf t})s_{\lambda'} ({\bf t'}) \ ,
\end{equation}
where the coefficients $g_{\lambda,\lambda'}$ can be presented
as certain determinants or, alternatively, just solve special
bilinear equations \cite{JM}.
We are interested in the semi-infinite Toda lattice, i.e., when the
index $n$ runs over the
set of nonnegative integers ${\mathbb Z}_{\ge0}$.
In the semi-infinite case, we have the restriction
$$ 
\ell (\lambda),\ell ({\lambda'})\le n
$$
of the summation range on the right hand side of (\ref{tauschurschur}). By
Theorem 1 of \cite{Tinit}, in this case
$$ 
g_{\lambda,{\lambda'}}(n)=\det
\left(G_{h_ih_j'}\right)_{i,j=1,\dots,n} \ ,
$$
$$ 
h_i=\lambda_i-i+n \ge0\ ,\quad h_i'=\lambda_i'-i+n\ge 0 \ ,
$$
and semi-infinite matrix $G$ is related to the factorization
problem $\Psi^+({\bf t},{\bf t'})=\Psi^-({\bf t},{\bf t'})G$,
where the semi-infinite matrices $\Psi^\pm$ are Baker-Akhiezer
functions \cite{UT}.

If all but finitely many terms in (\ref{tauschurschur}) vanish, it
gives a rational solution, in which case the formula may be best
known.
We are mainly interested in the case where (\ref{tauschurschur})
gives an infinite sum.

\paragraph*{Tau functions of hypergeometric type}

In the sequel, we shall also consider the {\em diagonal\/} case:
$g_{\lambda,\lambda'}(n)=0$ if $\lambda\ne\lambda'$. In this case,
generically (i.e., if $g_{{\bf0},\bf0}(n)\ne0$ ($\forall n$);
this restriction is not essential) we have
$g_{\lambda,\lambda'}(n)=\delta_{\lambda,{\lambda'}}r_\lambda(n)
g_{{\bf0},{\bf0}}(n)$, where the coefficients $r_\lambda(n)$ are
given in the following way:

Consider a function $r=r(n)$ of a single variable $n\in{\mathbb Z}$.
Given partition $\lambda$, we define
$$ 
r_\lambda(x)=\prod_{i,j\in\lambda}r(x+j-i)\,,\quad r_{\bf0}\equiv1\,.
$$
Namely, we place $r(x+j-i)$ at each node $(i,j)$ of diagram
$\lambda$, and multiply them over all nodes. The ``content'' $j-i$
of node $(i,j)$ vanishes on the main diagonal. For instance, if
$\lambda=(3,3,1)$, then (see the diagram in (\ref{YD331}))
$r_\lambda(x)=r(x+2)(r(x+1))^2(r(x))^2r(x-1)r(x-2)$.

We define the tau function of hypergeometric type as
\begin{equation} \label{exex'}
\tau_r(n,{\bf t},{\bf t'})=\sum_\lambda r_\lambda (n)
s_\lambda({\bf t})s_\lambda({\bf t'}) \ ,
\end{equation}
which solves the TL equation (\ref{Toda}), rewritten in the form
\begin{equation}\label{rToda'}
\partial_{t_1}\partial_{t'_1}\phi_n=
r(n)e^{\phi_{n-1}-\phi_n}- r(n+1)e^{\phi_n-\phi_{n+1}} \ ,
\quad \varphi_n=-\phi_n +\xi_n \ ,
\end{equation}
where we set $r(k)=e^{\xi_k-\xi_{k-1}}$ and
$$ 
\phi_n({\bf t},{\bf t'})=-\ln \frac{\tau_r(n+1,{\bf t},{\bf t'})}
{\tau_r(n,{\bf t},{\bf t'})}
$$
When $r(0)=0$, we obtain semi-infinite Toda lattice
(\ref{rToda'}).
{\em In what follows, when we consider semi-infinite Toda tau function
of the form $\tau_r$, we always assume $r(0)=0$.}

A simplest example yields a useful and simple relation (bosonic
(left hand side) and fermionic (right hand side) representations of the vacuum TL tau:
\begin{equation}\label{epp2}
\exp \sum_{m=1}^\infty m\gamma_m\gamma_m'= \sum_{\lambda \in P
}s_\lambda(\boldsymbol\gamma)s_\lambda(\boldsymbol\gamma') \ ,
\end{equation}
which is a generalized version of the {\em Cauchy-Littlewood
identity\/} \cite{Mac}.
Here we use notations
$\boldsymbol\gamma=(\gamma_1,\gamma_2,\dots)$ and
$\boldsymbol\gamma'=(\gamma_1',\gamma_2',\dots)$ in place of
${\bf t}$ and ${\bf t'}$, for future purposes.
(Since we did not find this identity in
literature, we presented a proof in \cite{Cadiz}).

\section{Discrete matrix models}
By discrete matrix model we mean models of random matrices, with
eigenvalues lying on a lattice, the type of the lattice is
different for different models and will be specified later in the
section "Associated discrete models and discrete one-matrix
models". We shall consider discrete counterparts of the following
models, where integrals over eigenvalues are replaced by sums.
Namely, we shall consider the following replacements.

Model of normal matrices:
\begin{equation}\label{d-nor}
\int_{{\mathbb C}^n}  |\Delta(z)|^2 \prod_{i=1}^n
e^{V_1(z_i)+V_2({\bar z}_i)}e^{V(z_i,{\bar z}_i)}d^2z_i  \to \sum
\cdots \sum |\Delta(z)|^2 \prod_{i=1}^n e^{V_1(z_i)+V_2({\bar
z}_i)}e^{V(z_i,{\bar z}_i)}=Z_1
\end{equation}
Model of two positive Hermitian matrices:
\begin{equation}\label{d-2-pos}
\int_{\mathbb R_+^{2n}} \Delta(x)\Delta(y) \prod_{i=1}^n
e^{V_1(x_i)+V_2({y}_i)}e^{V(x_i,{y}_i)}dx_idy_i \to \sum \cdots
\sum \Delta(x)\Delta(y) \prod_{i=1}^n e^{V_1(x_i)+V_2({
y}_i)}e^{V(x_i,{ y}_i)}=Z_2
\end{equation}
Let us notice that the well-known standard two-matrix model is
described by specific choice of the interaction term:
${V(x,{y})}=xy$.

Model of positive Hermitian matrices:
\begin{equation}\label{d-1-pos}
\int_{\mathbb R_+^{n}} \Delta(x)^2 \prod_{i=1}^n e^{V(x_i)}d x_i
\to \sum \cdots \sum \Delta(x)^2 \prod_{i=1}^n e^{V(x_i)}=Z_3
\end{equation}
Model of two unitary matrices
\begin{equation}\label{d-2-un}
\oint  \Delta(x)\Delta(y) \prod_{i=1}^n
e^{V_1(x_i)+V_2({y}_i)}e^{V(x_i,{y}_i)}\frac{dx_i dy_i}{x_iy_i}
\to \sum \cdots \sum \Delta(x)\Delta(y) \prod_{i=1}^n
e^{V_1(x_i)+V_2({y}_i)}e^{V(x_i,{y}_i)}=Z_4
\end{equation}
where, for the standard case, one takes
${V(x,{y})}=x^{-1}y^{-1}$.

Model of unitary matrices
\begin{equation}\label{d-1-un}
\oint  |\Delta(z)|^2 \prod_{i=1}^n e^{V_1(z_i)+V_2({\bar
z}_i)}e^{V(z_i,{\bar z}_i)}d^2z_i  \to \sum \cdots \sum
|\Delta(z)|^2 \prod_{i=1}^n e^{V_1(z_i)+V_2({\bar
z}_i)}e^{V(z_i,{\bar z}_i)}=Z_5
\end{equation}
Kontsevich-like model:
\begin{equation}\label{d-kon}
\int_{\mathbb R_+^{n}} \Delta(x) \prod_{i=1}^n
e^{x_iy_i}e^{V(x_i)}dx_i \to \sum \cdots \sum \Delta(x)
\prod_{i=1}^n e^{x_iy_i}e^{V(x_i)}=Z_6
\end{equation}

The multiply integrals in the left hand sides are integrals over
the eigenvalues of the related matrix models. The cases of the
generalized interaction term $V(x,y)$ in (\ref{d-2-pos}) and
(\ref{d-2-un}) should be considered separetely in order to be
considered as integrals over matices. We do it in section
"Two-matrix models with generalized interaction term" below.

\

What we are going to prove is that by choosing parameters ${\bf
t},{\bf t'}$ in a special way we obtain

\

\begin{equation}\label{}
Z_n(V;{\bf t},{\bf t'})= \mbox{patition function of discrete
matrix models }, Z_1,\dots,Z_6
\end{equation}

\section{Schur function expansion of the partition function}

In \cite{CRM}, \cite{Cadiz}, for the Gaussian interaction term
$V(z,{\bar z})=-|z|^2$, the following perturbation series was
obtained
\begin{equation}\label{singleschurexp}
Z_n(V;{\bf t},{\bf t'})=n!\sum_{\scriptstyle\lambda \in P\atop
\scriptstyle\ell(\lambda)\le n
}(n)_\lambda s_\lambda({\bf t})s_\lambda({\bf t'})
\ ,
\end{equation}
where $(n)_\lambda$ is defined in (\ref{Poch}).
At the present paper we generalize this result.

Given $V(z,{\bar z})$, let us introduce bi-moments $g_{km}$,
where $k,m=0,1,2,\dots$\,, as
\begin{equation}\label{K_{ij}}
g_{km}=\int z^k {\bar z}^m e^{V(z,{\bar z})} d^2z \ ,
\end{equation}
assuming the integral on the right hand side is well-defined.

\btheorem
The partition function of the model of the normal matrices
(\ref{normint-}) has the following perturbation series in the
coupling constants $\bf t$ and $\bf t'$:
\begin{equation}\label{doubleschurexp}
Z_n(V;{\bf t},{\bf t'})=n!\sum_{\scriptstyle \lambda,\lambda'\in P
\atop\scriptstyle \ell(\lambda),\ell(\lambda')\le n}
g_{\lambda,{\lambda'}}(n)s_\lambda({\bf t})s_{\lambda'}({\bf t'})
\ ,
\end{equation}
where
\begin{equation}\label{K=det}
g_{\lambda,{\lambda'}}(n)=\det
\left(g_{h_ih_j'}\right)_{i,j=1,\dots,n} \ ,
\end{equation}
\begin{equation}\label{h_i}
h_i=\lambda_i-i+n \ge0\ ,\quad h_i'={\lambda'}_i-i+n\ge 0
\end{equation}
\etheorem

Proof. The proof repeats one of the proofs of (\ref{singleschurexp}),
written down in \cite{CRM}. Taking $\gamma_m=t_m$, $\gamma_m'=(1/m)\Tr M^m$,
and applying (\ref{epp2}) to the term $e^{\Tr V_1}$, and applying it again
to the term $e^{\Tr V_2}$, with $\gamma_m=t_m'$,
$\gamma_m'=(1/m) \Tr (M^\dag)^m$, we obtain
\begin{eqnarray*}
\lefteqn{
Z_n(V;{\bf t},{\bf t'})
}\\
&=&
\sum_{\scriptstyle\lambda,\lambda'\in P\atop
\scriptstyle\ell(\lambda),\ell({\lambda'})\le n}s_\lambda({\bf t})
s_{\lambda'}({\bf t'})\int_{{\mathbb C}^n} \det\left(z_k^{\
h_i}\right)_{1\le k,i\le n} \det\left({\bar z_k}^{\
h_i'}\right)_{1\le k,i\le n} \prod_{i=1}^n
e^{V(z_i,{\bar z}_i)}d^2z_i \ ,
\end{eqnarray*}
where we used definition (\ref{detSc}) to get the
determinants in the integrands; these determinants are
parameterized by the sets $\{h_i\mid i=1,\dots,n\}$ and $\{h_i'\mid
i=1,\dots,n\}$. The rest of the proof is quite similar to the calculation
suggested by John Harnad for \cite{CRM}.

Given a pair of partitions (or, by (\ref{h_i}),
given a pair of sets $\{h_i,\},\{h_i'\}$), we develop each
determinant as a sum of $n!$ monomial terms: the first determinant
yields the sum of monomials over all permutations $\sigma$ of
numbers $1,\dots,n$: $\sum_{\sigma\in S_n}(-)^{{\rm sign}
(\sigma)}\prod_{i=1}^n z_i^{h_{\sigma(i)}}$, each monomial is
parameterized by the set of non-negative numbers $\{h_i\}$ and by
an element of the permutation group $\sigma$. Similarly, the second
determinant yields $\sum_\sigma(-)^{{\rm sign} (\sigma)}\prod_i \bar
z_i^{h_{\sigma(i)}'}$, each monomial is parameterized by the set
$\{h_i'\}$ and by $\sigma$. After integration, each pair of
monomials produces the product of $n$ integrals of type
(\ref{K_{ij}}). Notice the restriction
$\ell(\lambda)$, $\ell({\lambda'})\le n$, which follows from
Remark~\ref{Schurvanish}. Given a pair of sets $\{h_i\}$ and $\{h_i'\}$,
we gather all the
terms into determinant (\ref{K=det}) and obtain the Theorem.

\medskip

Note that the very first term of (\ref{doubleschurexp}) yields the formula
presented in \cite{Zabor}
$$ 
Z_n(V;0,0)=n!g_{{\bf0},\bf0}=n!\det\left(
\int z^{n-i}{\bar z}^{n-j}e^{V(z,\bar z)}d^2z\right)_{i,j=1,\dots,n}
$$

\begin{Corollary} Using Theorem 1 of \cite{Tinit}, we get another
proof of the fact that the partition function of the model of
normal matrices is a tau function of the (semi-infinite) TL hierarchy,
\cite{Zabor}, \cite{MWZ}.
\end{Corollary}

\paragraph*{Inverse moment problem}

One can ask what kind of potential $V(z,{\bar z})$ of MNM
(\ref{normint}) provides a given form of potentials in the
discrete matrix models we have obtained. We need to present the
following comment.
 \bremark\label{ana} Since the
moments $g_{mn}$ are the higher derivatives at the origin of the
Fourier transform of $e^V$, and since every formal power series is
the (formal) Taylor series at the origin of some $C^\infty$
function, {\em one can, in principle, regard the moments $g_{mn}$
as independent variables}. Having this fact in mind, in the future
we may write each bi-moment $g_{mn}$ of (\ref{K_{ij}}) in a form
analogous to $e^{V_1+V_2}e^V$ in (\ref{normint-}):
\begin{equation}\label{g-V}
g_{km}=e^{{\tilde V}_1(k)+{\tilde V}_2(m)}e^{{\tilde V}_{km}}
\ ,
\end{equation}
and assume certain kind of parameter dependence for $\tilde V_1(k)$
and $\tilde V_2(m)$.
\eremark

Note that, since the Taylor series of a $C^\infty$ function does not
determine the function uniquely, the moments $\{g_{mn}\}_{m,n\ge0}$ do
{\em not\/} determine $e^V$ uniquely.
Moreover, in popular function spaces like $L^2$ or Sobolev,
the set of all $e^V$'s whose bi-moments coincide with given $\{g_{mn}\}$
is not closed and may even be dense, so it is hard to give an explicit
method or formula to reconstruct one.

Nevertheless, among those $e^V$ there is {\em at most one} which
decays exponentially, i.e., $O(e^{-\varepsilon|z|})$ as $|z|\to\infty$,
since the Fourier transform of exponentially decreasing function is real
analytic, and hence is determined uniquely by its Taylor series.

Parameter dependence of $\{g_{km}\}$, like in (\ref{g-V}), may
often be introduced easily: Suppose, as a function
of $r\in\mathbb R$ and $w=e^{i\theta}\in\{|w|=1\}$,
$e^{V(z,\bar z)}=e^{V(rw,rw^{-1})}$ has analytic continuation
in the $w$-variable to ${\mathbb C}\setminus\{0\}$, and the
resulting function still is rapidly decreasing in $r$ for any
$w\in{\mathbb C}\setminus\{0\}$.
(This is slightly more general than the axial symmetric case.
It includes, e.g., polynomials in $z$ and $\bar z$ with
coefficients being rapidly decreasing functions in $|z|$.)
Let $e^{\tilde V_{km}}:=\int_{\mathbb C}z^k\bar z^me^{V(z,\bar z)}d^2z$.
Then for any $q_1$,~$q_2 >0$ we have
$$
q_1^{-1}q_2^{-1}\int_{\mathbb C}z^k\bar
z^me^{V(q_1^{-1}z,q_2^{-1}\bar z)}d^2z =q_1^kq_2^me^{\tilde
V_{km}}\,;
$$
hence, introducing two sets of parameters
$\btt=(\tilde t_1,\tilde t_2,\dots)$ and
$\btt'=(\tilde t_1',\tilde t_2',\dots)$, and
setting
\begin{equation}\label{Vtildet}
e^{V_{\btt,\btt'}(z,\bar z)}:=
e^{\xi(\btt,q_1^D)+\xi(\btt',q_2^{\bar D})}e^{V(z,\bar z)}=
\sum_{i,j\ge0}h_i(\btt)h_j(\btt')q_1^{-i}q_2^{-j}
e^{V(q_1^{-i}z,q_2^{-j}\bar z)}\,,
\end{equation}
where $D:=-z\partial/\partial z-1$ and
$\bar D:=-\bar z\partial/\partial\bar z-1$, at least as a formal
power series in $\btt$ and $\btt'$ we have
$$
\int_{\mathbb C}z^k\bar z^m e^{V_{\btt,\btt'}(z,\bar z)}d^2z
=e^{\xi(\btt,q_1^k)+\xi(\btt',q_2^m)}e^{\tilde V_{km}}\,.
$$
A simple diagram summarizes what we get:
\begin{equation}\label{momentdeform}
\begin{CD}
e^{V(z,\bar z)} @>{\rm deformation}>>
    e^{V_{\btt,\btt'}(z,\bar z)}\\
@V{\rm bimoments}VV     @VV{\rm bimoments}V \\
e^{\tilde V_{km}}@>{\rm deformation}>>
e^{\xi(\btt,q_1^k)+\xi(\btt',q_2^m)}e^{\tilde V_{km}}
\end{CD}
\end{equation}
It is more illuminating to put
$e^{\xi(\btt,q_1^D)+\xi(\btt',q_2^{\bar D})}e^{V(z,\bar z)}$
in place of $e^{V_{\btt,\btt'}(z,\bar z)}$ in this diagram,
but the construction (\ref{Vtildet}) may not work if $e^V$
is in more general form, or if the condition $q_1$,~$q_2>0$
is not satisfied.  Nevertheless, we can
rely on less explicit argument in Remark~\ref{ana} to assume
the existence of $e^{V_{\btt,\btt'}}$
subject to (\ref{momentdeform}).

\paragraph*{Axial-symmetric interaction term}
If we take
axial-symmetric interaction term (\ref{axial}), $V(z,{\bar
z})=V_{\rm axi}(|z|^2)$, then $g_{km}=\delta_{km}g_{mm}$ is a
diagonal matrix, which we parametrize
by a set of numbers $\xi_0$, $\xi_1$, $\xi_2,\dots$ as follows:
\begin{equation}\label{diag}
g_{mm}= \pi
\int_0^{+\infty} x^me^{V_{\rm axi}(x)} dx= e^{\xi_m-\xi_0} \ ,
\end{equation}
so that
$$ 
g_{\lambda,{\lambda'}}(n)=\delta_{\lambda,{\lambda'}}\prod_{i=1}^n
e^{\xi_{\lambda_i+n-i}-\xi_{n-i}}
$$
If both $\lambda$, $\lambda'$ are zero partitions, we have
$g_{{\bf0},{\bf0}}=\prod_{m=0}^{n-1}g_{mm}$.
Then we obtain
\begin{equation}\label{htf}
Z_n(V_{\rm axi};{\bf t},{\bf t'})=g_{{\bf0},{\bf0}}n!\sum_{\lambda \in P}
e^{\sum_{i=1}^\infty \xi_{n+\lambda_i-i}-\xi_{n-i}}s_\lambda({\bf t})
s_\lambda({\bf t'})
\end{equation}

Let us notice that a number of matrix integrals (including the
generalization of Itzykson-Zuber, Gross-Witten and Kontsevich
integrals, and various integrals over complex and rectangle
matrices) may be written in form of series (\ref{htf}), where the
variables $\xi$ are specialized via the choice of matrix model, see
\cite{1'}.

In the following sections, we consider the right hand side of
(\ref{doubleschurexp}) and of (\ref{htf}) as models
$Z_1,\dots,Z_6$, where sums over numbers $h=(h_1,\dots,h_n)$ and
$h'=(h'_1,\dots,h'_n)$ play the role of sums over eigenvalues.

\section{Specialization of coupling constants.
Associated discrete models and discrete one-matrix models}

Using  the Schur function development (\ref{doubleschurexp}), we
shall show that there exist associated discrete models of random
matrices, whose partition functions coincide with the partition
function of the model of the normal matrices.

We introduce the following notations:
\begin{equation}\label{choicetinfty'}
{\bf t}_\infty=(1,0,0,0,\dots) \ ,
\end{equation}
\begin{equation}\label{choicet(a)'}
{\bf t}(a,1)=\Bigl(\frac{a}{1},\frac{a}{2},\frac{a}{3},\dots\Bigr) \ ,
\end{equation}
\begin{equation}\label{choicetinftyq'}
{\bf t}(\infty,q)=(t_1(\infty,q),t_2(\infty,q),\dots),\quad
t_m(\infty,q)=\frac{1}{m(1-q^m)}\ ,\quad m=1,2,\dots\ ,
\end{equation}
\begin{equation}\label{choicet(a)q'}
{\bf t}(a,q)=(t_1(a,q),t_2(a,q),\dots)\ ,\quad
t_m(a,q)=\frac{1-(q^a)^{m}}{m(1-q^m)}\ ,\quad m=1,2,\dots
\end{equation}
These choice of times was also used in
\cite{pd22}--\cite{galipolli}.

Note that ${\bf t}(a,q)$ tends to ${\bf t}(\infty,q)$ (resp.\
${\bf t}(a,1)$) as $a\to\infty$ (resp.\ $q\to1$). As for ${\bf
t}_\infty$, if $f$ satisfies
$f(ct_1,c^2t_2,c^3t_3,\dots)=c^df(t_1,t_2,t_3,\dots)$ for some
$d\in {\mathbb Z}$, we have $\hbar^df({\bf t}(\infty,q))\to f({\bf
t}_\infty) $ as $\hbar:=\ln q \to0$.

\begin{Lemma} For a partition $\lambda=(\lambda_1,\lambda_2,\dots)$, let
$ 
h_i := n+\lambda_i-i$ $(1\le i \le n)$, where $n\ge\ell(\lambda))$.
Then
\begin{equation}\label{schurhook}
s_\lambda({\bf t}_\infty)=\frac{1}{H_\lambda}=\frac{
\Delta(h)}{\prod^n_{i=1}h_i!} \ ,
\end{equation}
\begin{equation}\label{schurhookt(a)}
s_\lambda({\bf t}(a,1))=\frac{(a)_\lambda}{H_\lambda}=\frac{
\Delta(h)}{\prod^n_{i= 1}h_i!} \prod_{i=1}^n
\frac{\Gamma(a-n+h_i+1)}{\Gamma(a-i+1)}\ ,
\end{equation}
\begin{equation}\label{schurhookq}
s_\lambda({\bf t}(\infty,q))
=\frac{q^{n(\lambda)}}{H_\lambda(q)}=\frac{\Delta(q^h)}
{\prod_{i=1}^n(q;q)_{h_i}}\ ,
\end{equation}
\begin{equation}\label{schurhookqa}
s_\lambda({\bf t}(a,q))=\frac{
q^{n(\lambda)}(q^a;q)_\lambda}{H_\lambda(q)}=\frac{\Delta(q^h)}
{\prod_{i=1}^n(q;q)_{h_i}}\prod_{i=1}^n (q^{a-i+1};q)_{h_i-n+i}
\ ,
\end{equation}
$$ 
\Delta(h):=\prod^n_{i<j}(h_i- h_j)\ ,\quad
\Delta(q^h):=\prod^n_{i<j}(q^{h_i}- q^{h_j}) \ ,
$$
where for $H_\lambda$, $H_\lambda(q)$, $n(\lambda)$,
$(a)_\lambda$ and $(q^a;q)_\lambda$ see respectively
(\ref{Hlambda}), (\ref{Hlambda(q)1}), (\ref{n(lambda)}),
(\ref{Poch}) and (\ref{Pochq}). Note that those quantities
(\ref{schurhook})--(\ref{schurhookqa}) are independent of the choice of
$n \ge \ell(\lambda)$.
\end{Lemma}

Relations (\ref{schurhook}) and (\ref{schurhookt(a)}) can be found
in \cite{Mac} (some notations are different) for
$n=\ell(\lambda)$, and we only need to add that if we have
(\ref{schurhook}) and other formulas in the Lemma for the value
$n=\ell(\lambda)$, then we have it for all $n>\ell(\lambda)$, as
seen by a straightforward calculation. First equalities in the
relations (\ref{schurhookq}) and (\ref{schurhookqa}) can be easily
obtained from well-known relations of \cite{Mac}. The second
equality in (\ref{schurhookqa}) follows from known formulae of
\cite{Mac}.

Using this Lemma, we consider the Schur function development of
the partition function of the model of the normal matrices
(\ref{doubleschurexp}). Choosing the higher times according to
(\ref{choicetinfty'}), to (\ref{choicet(a)'}), to
(\ref{choicetinftyq'}), and to (\ref{choicet(a)q'}), we obtain
the following four models (A), (B), (C) and (D), respectively:

\paragraph*{(A) Hermitian discrete matrix models $Z_2,Z_3$ with
positive eigenvalues}
Consider the case
\begin{equation}\label{t1tinfty}
{\bf t}=(t_1,0,0,\dots)=t_1{\bf t}_\infty\,,\quad
{\bf t'}=(t_1',0,0,\dots)=t_1'{\bf t}_\infty
\end{equation}
for model (\ref{normint}), and use $s_\lambda(t{\bf t}_\infty)
=t^{|\lambda|}s_\lambda({\bf t}_\infty)$. By Theorem 1
we have
$$ 
Z_n(V;{\bf t},{\bf t'})=C\int d\Omega(M)
e^{\Tr V(M,M^\dag)+t_1\Tr M+t_1'\Tr M^\dag}=n!
\sum_{\scriptstyle\lambda,\lambda'\atop
\scriptstyle\ell(\lambda),\ell({\lambda'})\le n}
t_1^{|\lambda|}(t_1')^{|{\lambda'}|}g_{\lambda{\lambda'}}(n)s_\lambda
({\bf t}_\infty)s_{\lambda'}({\bf t}_\infty)
$$
Note that this integral is a kind of Fourier transform.

For a pair of partitions $\lambda$, $\lambda'$, take any $n\in\mathbb Z$
such that $n\ge\ell(\lambda)$, $\ell(\lambda')$, and let
$h_i := n+\lambda_i-i$, $h_i':= n+\lambda'_i-i$, for $1\le i \le n$.
Then by using (\ref{schurhook}), the perturbation series for the partition
function of the model of the normal matrices becomes
\begin{eqnarray}
\lefteqn{
Z_n(V;{\bf t}_\infty,{\bf t}_\infty)
}\nonumber\\
&=& n!g_{{\bf0},{\bf0}}\sum_{h_1>\dots
>h_n\ge 0}\sum_{h'_1>\dots
> h'_n\ge 0} \Delta( h)\Delta( h')\prod_{i=1}^n e^{-\ln
\Gamma(h_i+1)-\ln \Gamma(h'_i+1)} \det
\left(g_{h_ih'_j}\right)_{i,j=1,\dots,n}
\nonumber\\
&=&\frac {g_{{\bf0},{\bf0}}}{n!}\sum_{h,h'\ge 0}
\Delta(
h)\Delta(h')\prod_{i=1}^n e^{-\ln \Gamma(h_i+1)-\ln
\Gamma(h'_i+1)} \det \left(g_{h_ih'_j}\right)_{i,j=1,\dots,n} \ ,
\label{discreteMMdet}
\end{eqnarray}
where we got the factor $1/n!$ instead of $n!$, as we switched
from the summation over the cones $h_1>\cdots>h_n\ge
0$, $h_1'>\cdots>h_n'\ge 0$ to the summations over all non-negative
integers $h_1,\dots,h_n$ and $h'_1,\dots,h'_n$, denoted by
$\sum_{h\ge 0}$ and by $\sum_{h'\ge 0}$ throughout the text.
According to (\ref{K_{ij}}), the factor $\det
\left(g_{h_ih'_j}\right)_{i,j=1,\dots,n}$ is defined by the
interaction term $V$ in the model of the normal matrices and
describes the interaction between the random matrices of the
discrete model.

Using the skew-symmetry of $\Delta( h)\Delta(h')$, one
can replace the determinant on the right hand side of
(\ref{discreteMMdet}) by the product of $n$ terms $g_{h_ih'_i}$,
$i=1,\dots,n$, getting $n!$ as a multiplier.

Recall that $|\lambda|=\sum_i h_i-(n^2-n)/2$. Finely, we find that
the perturbation series for the partition function of the model of
the normal matrices is equal to the following partition function
of a discrete model:
\begin{equation}\label{discreteMM}
Z_n(V;{\bf t},{\bf t'})=
(t_1t_1')^{\frac{n-n^2}{2}}g_{{\bf0},{\bf0}}\sum_{h,h'\ge
0} \Delta(h)\Delta( h')\prod_{i=1}^n e^{-\ln \Gamma(h_i+1)-\ln
\Gamma(h'_i+1)+h_i\ln t_1+h_i'\ln t_1' +\ln g_{h_ih'_i}} \ ,
\end{equation}
where each sum ranges over the $n$-tuples of all non-negative integers,
$h_1,\dots,h_n$ and $h_1',\dots,h_n'$, respectively; the times
${\bf t},{\bf t'}$ in the left hand side are chosen according to
(\ref{t1tinfty}), and we remind that
$$ 
g_{km}=\int_{\mathbb C}z^k{\bar z}^m e^{V(z,{\bar z})} d^2z\ ,\quad
\Delta(h)=\prod^n_{i<j}(h_i-h_j)\ ,\quad
\Delta(h')=\prod^n_{i<j}(h'_i-h'_j)
$$
We notice that the first
non-trivial term in sum (\ref{discreteMM}), corresponding to $h_i=h'_i=n-i$,
is equal to one.

In case $g_{km}=f(km)e^{{\tilde V}_1(k)+{\tilde V}_2(m)}$ (see
Remark~\ref{ana}) the sum (\ref{discreteMM}) is a discrete
analogue of the model of two Hermitian random matrices with
non-negative eigenvalues with generalized interaction term $Z_2$,
see (\ref{eigengenmulti}) below. In case $f(km)=e^{-km}$ we obtain
a discrete analogue of standard model of two Hermitian random
matrices.

If we take {\em axial-symmetric} interaction term (\ref{htf}), we
obtain the partition function of the discrete one-matrix model of
non-negative matrices $Z_3$:
\begin{equation}\label{1MMdiscreteaxial-3}
Z_n(V_{\rm axi};{\bf t}_\infty,{\bf t}_\infty)= (t_1t_1')^{\frac{n-n^2}{2}}
g_{{\bf0},{\bf0}}
\sum_{h\ge 0} \Delta(h)^2\prod_{i=1}^n e^{-2\ln h_i!+h_i\ln(t_1
t_1')+\xi_{h_i}-\xi_{n-i}}
\end{equation}

\paragraph*{(B)} We obtain different versions of models $Z_2,Z_3$ if we choose
\begin{equation}\label{t(a)t'(a')}
{\bf t}=(xa,x^2a,x^3a,\dots)\ , \quad {\bf t'}=(ya',y^2a',y^3a',\dots)
\end{equation}
Then, (as done in \cite{1'}) one can see that (\ref{normint})
takes the form
$$ 
Z_n(V;{\bf t},{\bf t'})=C\int d\Omega(M) e^{\Tr
V(M,M^\dag)}\det(I_n-xM)^{-a}\det(I_n-yM^\dag)^{-a'}
$$
Using results of \cite{Mac} and the definition of $h_i$, we find
that, for ${\bf t}$ as in (\ref{t(a)t'(a')}),
$$ 
s_\lambda({\bf t})=
\frac{x^{|\lambda|}}{H_\lambda}\prod_{i=1}^n
\frac{\Gamma(a-n+h_i+1)}{\Gamma(a-i+1)}
$$
Now, for ${\bf t},{\bf t'}$ of form (\ref{t(a)t'(a')}), we get a
partition function with the potential terms slightly different of
the similar terms in (\ref{discreteMM}):
$$ 
Z_n(V;{\bf t},{\bf t'})=(xy)^{\frac{n-n^2}{2}}g_{{\bf0},{\bf0}}\sum_{h,h'\ge 0}
\Delta(h)\Delta( h')\prod_{i=1}^n e^{\ln
\frac{\Gamma(a-n+h_i+1)}{\Gamma(a-i+1)\Gamma(h_i+1)}+\ln
\frac{\Gamma(a'-n+h'_i+1)}{\Gamma(a'-i+1)\Gamma(h'_i+1)}+h_i\ln
x+h'_i\ln y+\ln g_{h_ih'_i}}
$$

For axial-symmetric case (\ref{axial}),
we obtain the following partition function of the discrete model of random
non-negative matrices:
\begin{equation}\label{discreteMMBaxial}
Z_n(V_{\rm axi};{\bf t},{\bf t'})=(xy)^{\frac{n-n^2}{2}}g_{{\bf0},{\bf0}}
\sum_{h\ge0}\Delta(h)^2\prod_{i=1}^n
e^{\ln\frac{\Gamma(a-n+h_i+1)\Gamma(a'-n+h_i+1)}{\Gamma(a-i+1)
\Gamma(a'-i+1)}-2(h_i!)^2 +h_i\ln (xy)+\xi_{h_i}-\xi_{n-i}} \ ,
\end{equation}
where variables $T_k$ are related to the axial-symmetric $V$ via
(\ref{diag}).

Another example: Take the axial symmetric Gauss interaction
term $V_{\rm axi}=-\Tr (MM^\dag)$. Then
\begin{eqnarray*}
Z_n(V_{\rm axi};{\bf t}(a,1),{\bf t}_\infty) &=&C\int
d\Omega(M)e^{-\Tr (MM^\dag)+t_1\Tr M}\det(I_n-yM^\dag)^{-a} =
g_{{\bf0},{\bf0}}\sum_\lambda
(yt_1)^{|\lambda|}\frac{(n)_\lambda(a)_\lambda}{H_\lambda^2}
\\
&=& (yt_1)^{\frac{n-n^2}{2}}
g_{{\bf0},{\bf0}}\sum_{h\ge 0} \Delta(h)^2\prod_{k=1}^n
k!e^{\ln \frac{\Gamma(a-n+h_k+1)}{\Gamma(a-i+1)}-2\ln h_k! +h_k\ln
(yt_1) }
\end{eqnarray*}

\paragraph*{(C) Discrete models related to unitary, $Z_4,Z_5$,
and to normal matrices $Z_1$} Choosing the times as in
(\ref{choicetinftyq'}), we obtain that the partition function of
the MNM (\ref{normint}) takes a form
$$ 
Z_n(V;{\bf t},{\bf t'})=C\int e^{\Tr V(M,M^\dag)}\frac{
d\Omega(M)} {\det(xM;q_1)_\infty\det(yM^\dag;q_2)_\infty } \ ,
$$
where we use the notation (\ref{Pochq}) for $m=\infty$, and where
$$ 
{\bf t}= (x\frac{1}{1-q_1},\frac{x^2}{2}\frac{1}{1-q_1^2},
\frac{x^3}{3}\frac{1}{1-q_1^3}, \dots) \ , \quad {\bf t'}=
(y\frac{1}{1-q_2},\frac{y^2}{2}\frac{1}{1-q_2^2},
\frac{y^3}{3}\frac{1}{1-q_2^3}, \dots)
$$

Using the arguments similar to the previous cases, we obtain the
following discrete model
$$ 
Z_n(V;{\bf t},{\bf t'})= (xy)^{\frac{n-n^2}{2}}g_{{\bf0},{\bf0}} \sum_{h,h'\ge 0} \Delta(q_1^h)\Delta(q_2^{h'})\prod_{i=1}^n
e^{-\ln (q_1;q_1)_{h_i}-\ln (q_2;q_2)_{h_i'}+h_i\ln x+h_i'\ln y
+\ln g_{h_ih'_i}}
$$

Let us take $q_1=q_2=q$ and $|q|=1$ (then we set
$q=e^{\sqrt{-1}\hbar}$).

In case $g_{km}=f(q^kq^m)e^{{\tilde V}_1(q^{-k})+{\tilde
V}_2(q^{-m})}$ (see Remark~\ref{ana}) the sum (\ref{discreteMM})
is a discrete analogue of the model of two unitary random matrices
with generalized interaction term $Z_4$, see (\ref{eigengenmulti})
below. In case $f(q^kq^m)=e^{q^kq^m}$ we obtain a discrete
analogue of standard model of two unitary random matrices.

For axial-symmetric interaction term $V$ (\ref{axial}) in case
$q_1={\bar q}_2=q$ and using parametrization (\ref{diag}), we
obtain the discrete model one-matrix model:
\begin{equation}\label{discrete1MMC}
Z_n(V_{\rm axi};{\bf t},{\bf t'})=
(xy)^{\frac{n-n^2}{2}}g_{{\bf0},{\bf0}}
\sum_{h\ge 0}
|\Delta(q^h)|^2\prod_{i=1}^n e^{-2\ln (q;q)_{h_i}+h_i\ln (xy)
+\xi_{h_i}-\xi_{n-i}} \ ,
\end{equation}
which, in the case where $|q|=1$, can be interpreted as model of
unitary matrices $Z_5$, and in the case where $|q|<1$, gives a
discrete version of the model of the normal random matrices $Z_1$,
where instead of the integration over the complex plane, we have
summation over the spiral $\{q^h\mid h=0,1,\dots\}$.

\paragraph*{(D)} This case is rather close to the previous one.
The choice of times as
$$ 
{\bf t}=
\biggl(x\frac{1-q_1^a}{1-q_1},\frac{x^2}{2}\frac{1-q_1^{2a}}{1-q_1^2},
\frac{x^3}{3}\frac{1-q_1^{3a}}{1-q_1^3}, \dots\biggr) \ ,
\quad{\bf t'}= \biggl(y\frac{1-q_2^{a'}}{1-q_2},\frac{y^2}{2}
\frac{1-q_2^{2a'}}{1-q_2^2},
\frac{y^3}{3}\frac{1-q_2^{3a'}}{1-q_2^3}, \dots\biggr)
$$
in the MNM model (\ref{normint}) yields the integral
$$ 
Z_n(V;{\bf t},{\bf t'})=C\int d\Omega(M)
e^{\Tr V(M,M^\dag)}\det
\frac{(xq_1^aM;q_1)_\infty(yq_2^{a'}M^\dag;q_2)_\infty}
{(xM;q_1)_\infty(yM^\dag;q_2)_\infty}
$$
This matrix integral have the perturbation series, described by
the following discrete model:
$$ 
Z_n(V;{\bf t},{\bf t'})= (xy)^{\frac{n-n^2}{2}}g_{{\bf0},{\bf0}}
\sum_{h,h'\ge 0} \Delta(q_1^h)\Delta(q_2^{h'})\prod_{i=1}^n
e^{\ln\frac{(q_1^{a-i+1};q_1)_{h_i-n+1}(q_2^{a'-i+1};q_2)_{h_i'-n+1}}{
(q_1;q_1)_{h_i}(q_2;q_2)_{h_i'}}+h_i\ln x+h_i'\ln y +\ln
g_{h_ih'_i}}
$$
For axial-symmetric case (\ref{axial}), and for the case $q_1={\bar q}_2=q$,
we have the model
\begin{equation}\label{discrete1MMD}
Z_n(V_{\rm axi};{\bf t},{\bf t'})=
(xy)^{\frac{n-n^2}{2}}g_{{\bf0},{\bf0}} \sum_{h\ge
0} (\Delta(q^h))^2\prod_{i=1}^n
e^{\ln\frac{(q^{a-i+1};q)_{h_i-n+1}(q^{a'-i+1};q)_{h_i-n+1}}
{(q;q)_{h_i}^2}+h_i\ln (xy)
+\xi_{h_i}-\xi_{n-i}} \ ,
\end{equation}
where the right hand side in the case $|q|=1$ can be interpreted
as a discrete one-matrix model $Z_5$, and for $q_{1,2}=\rho
e^{\pm\sqrt{-1}\hbar}$ it is a version of discrete MNM model
$Z_1$.

\medskip

\paragraph*{(E) Kontsevich-like discrete model $Z_6$. }

To get model (\ref{d-kon})  we consider the axial symmetric case.

First, we choose the variables ${\bf t'}={\bf t'}(y^{(n)})$. Then,
we specialize $\bf t$ according to any of (\ref{choicetinfty'}),
(\ref{choicet(a)'}), (\ref{choicetinftyq'}), or
(\ref{choicet(a)q'}).

Let us consider only the case when $\bf t$ is specialized by
(\ref{choicet(a)q'}) (which is the most general case among
specializations (\ref{choicetinfty'}),  (\ref{choicet(a)'}),
(\ref{choicetinftyq'}), (\ref{choicet(a)q'})). Then, using
(\ref{htf}), using definition (\ref{detSc}) of the Schur function,
using (\ref{schurhookqa}) and using antisymmetry of the
Vandermonde determinant, we can rewrite (\ref{doubleschurexp}) in
the form

\begin{equation}\label{ADMII}
Z_n(V_{\rm axi},{\bf t}(a,q),{\bf t}(y^{(n)}))\Delta(y)=
g_{{\bf0},{\bf0}}n! \cdot\sum_{h\ge0} \Delta(q^h)
\frac{\prod_{i=1}^n (q^{a-i+1};q)_{h_i-n+i}}
{\prod_{i=1}^n(q;q)_{h_i}} e^{\sum_{i=1}^\infty
(\xi_{h_i}-\xi_{n-i}+h_i\ln y_i)}
\end{equation}

By changing $\log y_i \to  y_i$ we obtain $Z_6$.

\

We mark that the relations between continues and discrete
models of type, different of presented here, were considered
in the papers \cite{Kos1}, \cite{Kos2}.

\section{Dual Schur functions.
Associated discrete models and discrete matrix models}

In this section, we shall specify the sets of coupling constants
${\bf t}$ and ${\bf t'}$ in (\ref{doubleschurexp}) in a different
way, via partitions, say, $\lambda^*$ and $\lambda'^*$. To do it,
we consider the Schur function $\underline s_\lambda(x)$ in (\ref{detSc})
as the function of a partition $\lambda$. We show that if
$x_i=q^{h_i^*}$, where $h_i^*$ are integers, one can interchange
the roles of $x$ and $\lambda$. Using this fact we differently
relate series (\ref{doubleschurexp}) to discrete analogs of
partition functions of matrix models.

Given partitions $\lambda$ and $\lambda^*$, whose lengths do not exceed
a given number $n$, we let
\begin{equation}\label{h-lambda}
h_i=\lambda_i-i+n\ ,\quad h_i^*=\lambda^*_i-i+n \ ,
\end{equation}
so that we have
$h_1>h_2>h_3>\cdots>h_n\ge0$, and
$h_1^*>h_2^*>h_3^*>\cdots>h_n^*\ge0$.
Let
\begin{equation}\label{x*-h}
x_i^*=q^{h_i},\quad x_i=q^{h_i^*},\quad i=1,\dots,n\ ,
\end{equation}
where $q$ is a given complex number.

By definition (\ref{detSc}) of the Schur function,
we obtain the following Lemma
\begin{Lemma} For any $q$, let $x=(x_1,\dots,x_n)$ and
$x^*=(x_1^*,\dots,x_n^*)$ be as in (\ref{x*-h}). Then we have
\begin{equation}\label{Schurf-duality}
\Delta(x)\underline s_\lambda(x)=\Delta(x^*)\underline s_{\lambda^*}(x^*)
\end{equation}
\end{Lemma}

Thus, given $\underline s_\lambda(x)$, with $x$ as described
above, we obtain the dual Schur function $s_{\lambda^*}(x^*)$,
where the roles of $h$ and $h^*$ are interchanged. Letting $q\to
1$ in (\ref{Schurf-duality}), we get $\Delta(h^*)s_\lambda({\bf
t}(n,1))=\Delta(h) s_{\lambda^*}({\bf t}(n,1))$ in accordance with
(\ref{schurhookqa}).

We remark that the case (\ref{choicetinftyq'}) results from
(\ref{x*-h}) for $h_i^*=n-i$ and $n \to\infty$.

\medskip

Given complex $q_1$ and $q_2$ and given partitions $\lambda^*$
and $\lambda'^*$, we now consider
\begin{equation}\label{discretexy}
x_i=q_1^{h_i^*},\quad y_i=q_2^{{h_i'}^*} \ ,
\end{equation}
where
\begin{equation}\label{h*h'*-lambda*-nu*}
h_i^*=\lambda_i^*-i+n\ge0\ ,\quad
{h_i'}^*={\lambda'_i}^*-i+n\ge 0 \ ,
\end{equation}
putting
$$ 
x_i^*=q_1^{h_i},\quad y_i^*=q_2^{h_i'} \ ,
$$
$$ 
h_i=\lambda_i-i+n \ge0\ ,\quad h_i'={\lambda'}_i-i+n\ge 0
$$

We choose ${\bf t}(x)=(t_1,t_2,\dots)$, ${\bf t'}(y)=(t_1',t_2',\dots)$ as
\begin{equation}\label{t(x)t*(y)}
t_m=\frac 1m\sum_{i=1}^n q_1^{mh_i^*}\ ,\quad t_m'=\frac 1m
\sum_{i=1}^n q_2^{{mh_i'}^*}\ ,\quad m=1,2,\dots
\end{equation}
For the case when $x_i=q^{n-i}$, $i=1,\dots,n $, that is
$\lambda^*=0$, we shall write
\begin{equation}\label{choicetinftyq}
{\bf t}(n,q) =(t_1(n,q),t_2(n,q),\dots)\ ,\quad t_m(n,q)=\frac
1m\sum_{i=1}^n q^{m(n-i)}\ ,\quad m=1,2,\dots
\end{equation}

Then, using Lemma 1, we obtain for (\ref{doubleschurexp})
$$ 
Z_n(V;{\bf t}(x),{\bf t'}(y))\Delta(x)\Delta(y)=n!\sum_{h,h'}\det
\left(g_{h_ih_j'}\right)_{i,j=1,\dots
,n}s_{\lambda^*}(q_1^h)s_{\lambda'^*}(q_2^{h'})\Delta(q_1^h)\Delta(q_2^{h'})
\ ,
$$
where the sums range over the cones
$$ 
\infty > h_1>\cdots > h_n\ge0\ ,\quad \infty > h_1'
>\cdots > h_n'\ge 0 \ ,
$$

For the choice of ${\bf t}(x)$ and ${\bf t'}(y)$, given by
partitions $\lambda^*$ and $\lambda'^*$ via
(\ref{t(x)t*(y)}) and (\ref{discretexy}), we shall use the
notation
\begin{equation}\label{Z(lambda*nu*)}
Z_n(V;\lambda^*,\lambda'^*;q_1,q_2):=Z_n(V;{\bf t}(x),
{\bf t'}(y))\Delta(x)\Delta(y) \ ,
\end{equation}
where the right hand side is defined by (\ref{doubleschurexp}). The case
$\lambda^*=\lambda'^*=0$ is related to (\ref{choicetinftyq}).

Let us note that, in the case $n \to \infty$, the choice of the coupling constants in form
(\ref{t(x)t*(y)}) is rather general.

Using skew-symmetry properties of the factors, and changing the
sum over partitions to the sum over all non-negative integers, we
get
$$ 
Z_n(V;\lambda^*,\lambda'^*;q_1,q_2)=n!g_{{\bf0},{\bf0}}\sum_{h,h'\ge 0}
\frac{1}{(n!)^2}\Delta(q_1^h)\Delta(q_2^{h'})
s_{\lambda^*}(q_1^h)s_{\lambda'^*}(q_2^{h'})
\prod_{i=1}^n\left(g_{h_ih_i'}\right) \ ,
$$
where we use notation (\ref{Z(lambda*nu*)}), and where the sums
range over all the $n$-tuples of non-negative integers $h_1,\dots,h_n$ and
$h_1',\dots,h_n'$.
(We kept $1/(n!)^2$ only to explain that the $(n!)^2$ in the
denominator is obtained when we change the sum over cones to the
sum over all non-negative integers).

\paragraph*{Associated discrete model}
Let us introduce the following partition function
\begin{equation}\label{discrete2MMq1q2'}
Z^{\rm discr}_n(\tilde V;\btt,
\btt';q_1,q_2)=\frac{1}{n!} \sum_{h,h'\ge 0}
\Delta(q_1^h)\Delta(q_2^{h'}) \prod_{i=1}^n e^{\xi(\btt,
q_1^{h_i})+\xi( {\btt'},q_2^{h_i'})}e^{\tilde
V_{h_ih_i'}}
\end{equation}
This sum can be viewed as a partition function of different
models, which are discrete analogue of the models of random
matrices, the choice of the model of random matrices depending on
the specialization of the complex numbers $q_1$ and $q_2$. For
instance, for $q_1=\rho e^{\sqrt{-1}\hbar}$ and $q_2=\rho
e^{-\sqrt{-1}\hbar}$, where $\rho \neq 1$, sum
(\ref{discrete2MMq1q2'}) can be considered as a discrete version
of (\ref{normint-}).

Considering the perturbation series in coupling constants
$\btt$, $\btt'$ of (\ref{discrete2MMq1q2'}) we can
repeat the same trick we used to get Theorem 1. We obtain the
series of type (\ref{doubleschurexp}).
\begin{theorem}
Let
\begin{equation}\label{K_{ij}-discr}
{\tilde g}_{km}=A\sum_{h,h'\ge 0} \ \prod_{i=1}^n q_1^{k h_i}
q_2^{m{h'}_i}e^{{\tilde V}_{h_i{h'}_i}}\ (\neq\infty)\ ,\quad
k,m=0,1,2,\dots \ ,
\end{equation}
where the sum range over all non-negative numbers $h_1,\dots,h_n$ and
$h'_1,\dots,h'_n$, and $q_1$ and $q_2$ are given complex numbers.
The perturbation series in the coupling constants $\btt$ and
$\btt'$ for partition function of the model
(\ref{discrete2MMq1q2'}) can be presented as follows:
$$ 
Z^{\rm discr}_n(\tilde V;\btt,\btt';q_1,q_2)=
\sum_{\scriptstyle\lambda,{\lambda'}\in P \atop
\scriptstyle\ell(\lambda),\ell({\lambda'})\le n}{\tilde
g}_{\lambda,{\lambda'}}(n)s_\lambda(\btt)
s_{\lambda'}(\btt') \ ,
$$
where the sums range over all non-negative numbers $h_1,\dots,h_n$ and
$h'_1,\dots,h'_n$, and where
$$ 
{\tilde g}_{\lambda,{\lambda'}}(n)=\det \left({\tilde g}_{{\tilde
h}_k {\tilde h}_m'}\right)_{k,m=1,\dots,n} \ ,
$$
$$ 
{\tilde h}_i=\lambda_i-i+n\ge0\ ,\quad {\tilde
h}_i'={\lambda'}_i-i+n\ge 0
$$
This shows that sum (\ref{discrete2MMq1q2'}) is a TL tau
function, where $n,\btt,\btt'$ play the role
of higher times.
\end{theorem}
\begin{remark}
The sums (\ref{discreteMMdet}), (\ref{discrete2MMq1q2'}) can be
studied by the method of (discrete) orthogonal
polynomials (about the discrete version see, for instance,
\cite{BB}), which links these sums with tau functions of integrable
hierarchies, as done in \cite{GMMMO} for the continuous case (namely,
when we have integrals, instead of sums).
\end{remark}

Consider the differential operator
$s_{\lambda^*}({\tilde\partial})$, which is evaluated as the Schur function (\ref{Schurt}), where each variable $t_m$ is replaced by
the differential operator
${\tilde\partial}_m=(1/m)(\partial/\partial\tilde t_m)$, and where the partition
$\lambda^*$ is
related to the variables $x$ by
(\ref{h*h'*-lambda*-nu*}) and (\ref{discretexy}).
Similarly, consider $s_{\lambda'^*}({\tilde\partial}')$, which
is the Schur function (\ref{Schurt}), where each $t_m$ is replaced
by ${\tilde\partial}_m'=(1/m)(\partial/\partial\tilde t_m')$,
and where the partition $\lambda'^*$ is related to the
variables $y$ by (\ref{h*h'*-lambda*-nu*}) and by
(\ref{discretexy}).

By Lemma 2, we obtain
\begin{theorem} Let $g_{km}=e^{\tilde V_{km}}$.
Suppose conditions (\ref{K_{ij}}) and (\ref{K_{ij}-discr}) of
Theorems 1 and 2 are satisfied.
Let $a_n:=g_{{\bf0},\bf0}(n)/\tilde g_{{\bf0},\bf0}(n)$.
In the notation (\ref{Z(lambda*nu*)}),
we have
\begin{equation}\label{discrete2MMq1q2}
Z_n(V;0,0;q_1,q_2)= a_nZ^{\rm discr}_n(\tilde V;
\btt,\btt';q_1,q_2)|_{\btt,\btt'=0} \ ,
\end{equation}
\begin{equation}\label{Schur-diff}
Z_n(V;\lambda^*,\lambda'^*;q_1,q_2)=
a_ns_{\lambda^*}({\tilde\partial})s_{\lambda'^*}({\tilde\partial}')
\cdot Z^{\rm discr}_n(\tilde V;\btt,\btt';
q_1,q_2)|_{\btt,\btt'=0}
\end{equation}
If, moreover, the potential $V$ is deformed to
$V_{\btt,\btt'}$ subject to the diagram
(\ref{momentdeform}), then we can remove the restriction to
$\btt$,~$\btt'=0$ in (\ref{discrete2MMq1q2}) and
(\ref{Schur-diff}):
\begin{equation}\label{discrete2MMq1q2withTildeT}
Z_n(V_{\btt,\btt'};0,0;q_1,q_2)= a_nZ^{\rm discr}_n(\tilde V;
\btt,\btt';q_1,q_2)\ ,
\end{equation}
\begin{equation}\label{Schur-diffwithTildeT}
Z_n(V_{\btt,\btt'};\lambda^*,\lambda'^*;q_1,q_2)=
a_ns_{\lambda^*}({\tilde\partial})s_{\lambda'^*}({\tilde\partial}')
\cdot Z^{\rm discr}_n(\tilde V;\btt,\btt';
q_1,q_2)
\end{equation}
\end{theorem}
We hope that the problem of averaging of the
partition function (\ref{normint-}) over initial data, namely,
over ${\bf t}$ and ${\bf t'}$, can be related to the problem of
random partitions \cite{Olshanski}, $\lambda^*$ and $\lambda'^*$.


\paragraph*{Axial-symmetric case (\ref{axial})} In this case $h=h'$.
We denote $\tilde V(h_i):=\tilde V_{h_ih_i}$.
We have
\begin{equation}\label{discrete-axial}
Z^{\rm discr}_n(\tilde V;\btt,\btt';
q_1,q_2)=\frac{\tilde g_{{\bf0},{\bf0}}}{n!} \sum_{h\ge 0}
\Delta(q_1^h)\Delta(q_2^h) \prod_{i=1}^n e^{\xi(\btt,
q_1^{h_i})+\xi( {\btt'},q_2^{h_i})}e^{\tilde
V(h_i)} \ ,
\end{equation}
where the sum ranges over all non-negative integer $h_1,\dots,h_n$, and
\begin{equation}\label{Schur-diff-axial}
Z_n(V_{\rm axi};\lambda^*,\lambda'^*;q_1,q_2)=
a_ns_{\lambda^*}({\tilde\partial})s_{\lambda'^*}({\tilde\partial}')
\cdot Z^{\rm discr}_n(\tilde V;\btt,\btt';
q_1,q_2)|_{\btt,\btt'=0} \ ,
\end{equation}
where ${\bf t}(x)$ and ${\bf t'}(y)$ are described by the
partitions $\lambda^*$ and ${\lambda'}^*$ as described by
(\ref{t(x)t*(y)}) and (\ref{h*h'*-lambda*-nu*}). For
$\lambda^*={\lambda'}^*=0$, we have
$s_{\lambda^*}({\tilde\partial})=s_{\lambda'^*}({\tilde\partial}')=1$,
therefore, the partition functions of the continues
and discrete models coincide.

We have different models for different $q_1$ and $q_2$:

(A) If $q_1={\bar q}_2=e^{\sqrt{-1}\hbar}$ and all ${\tilde
V}_{mm}=0$, then, ({\ref{discrete-axial}), where
$\tilde p=\tilde p'=0$, is a discrete version of the partition
function of unitary one-matrix model
\cite{ZKMMO}, \cite{galipolli}, that is
\begin{equation}\label{D3}
\int_{U(n)}e^{ \sum_i t_i\Tr U^i +\sum_i
t'_i\Tr U^{-i}}d_*U=\sum_{\lambda:\ell(\lambda)\le n}
s_\lambda({\bf t}) s_\lambda ({\bf t'})=\tau_r(n,{\bf t},{\bf t'})
\ ,
\end{equation}
where we set ${\bf t}=\btt$, ${\bf t'}=\btt'$.
In (\ref{D3}), the $r$ is the following step function:
$r(k)=1$ if $k>0$, $r(k)=0$ if $k\le 0$.

(B) Now, $q_1=q=\rho e^{\sqrt{-1}\hbar}$ is a complex conjugate
to $q_2$, and $|\rho|$,~$|\hbar|\ll 1$. Let each $t_m'={\bar t}_m$ is
a complex conjugate to $t_m$, that is, by (\ref{t(x)t*(y)}), we
have $h_i^*={h_i'}^*$ for each $i$. We obtain
$$ 
Z_n(V_{\rm axi};0,0;q,{\bar q})=\frac{a_n}{n!}\sum_{h\ge 0}
\left|\Delta(q^h)\right|^2\prod_{i=1}^n e^{\tilde
V(h_i)} \ ,
$$
which is a discrete analogue of the axial-symmetric model of the
normal matrices (\ref{normint-}) and (\ref{axial}), and
$$ 
Z_n(V_{\rm axi};\lambda^*,\lambda'^*;q_1,q_2)=\frac{a_n}{n!}
s_{\lambda^*}({\tilde\partial})s_{\lambda'^*}({\tilde\partial}')
\cdot\sum_{h\ge 0} \left|\Delta(q^h)\right|^2\prod_{i=1}^n
e^{\xi(\btt,
q^{h_i})+\xi( {\btt'},{\bar q}^{h_i})}
e^{ \tilde
V(h_i)}|_{\btt,\btt'=0}
$$


\section{Discrete models as soliton solutions}

Here we consider the axial-symmetric case.
In Frobenius notations (see \cite{Mac}}), $\lambda=(\alpha|\beta)$,
we have
$$
e^{\sum_{i=1}^n
(\xi_{h_i}-\xi_{n-i})}= e^{\sum_{i=1}^k
(\xi_{n+\alpha_i}-\xi_{n-\beta_i-1})}
$$
and the series (\ref{htf}) may be restated as
\begin{equation}
Z_n(V_{\rm axi};{\bf t},{\bf t'})=n!g_{{\bf0},{\bf0}}
\left(
1+\sum_{k=1}^\infty\sum_{\scriptstyle\alpha_1>\cdots>\alpha_k\ge0\atop
\scriptstyle n>\beta_1>\cdots>\beta_k\ge0 }^\infty e^{\sum_{i=1}^k
(\xi_{n+\alpha_i}-\xi_{n-\beta_i-1})}s_{(\alpha|\beta)}({\bf t})
s_{(\alpha|\beta)}({\bf t'})\right)
\end{equation}
According to (\cite{hypsol}), when we specialize ${\bf t}$ by
(\ref{choicetinfty'})--(\ref{choicet(a)q'}), we obtain soliton tau
functions of certain (dual) TL equation.

For instance, if we take ${\bf t}={\bf t}_\infty$, we obtain
$$
(n!g_{{\bf0},{\bf0}})^{-1}Z_n(V_{\rm axi};{\bf t}_\infty,{\bf t'})=
$$
$$
1+\sum_{k=1}^\infty\sum_{\scriptstyle\alpha_1>\cdots>\alpha_k\ge0 \atop
\scriptstyle n>\beta_1>\cdots>\beta_k\ge0 }^\infty
\frac {\prod^k_{i<j}(\alpha_i-\alpha_j)(\beta_i-\beta_j)}
{\prod_{i,j=1}^k(\alpha_i+\beta_j+1)}
\frac {s_{(\alpha|\beta)}({\bf t'})}{\prod_{i=1}^k
\alpha_i!\prod_{i=1}^k \beta_i!}
e^{\sum_{i=1}^k
(\xi_{n+\alpha_i}-\xi_{n-\beta_i-1})}
$$
Then, if we take the collection $\xi_0$, $\xi_1$, $\xi_2,\dots$ by
$\xi_m:=\xi(\btt,m+a)-\xi({\btt'},(m+a)^{-1})
=\sum_{k=1}^\infty ((m+a)^k \tilde t_k-(m+a)^{-k}\tilde t_k')$,
which, for any choice of ${\bf t'}$, is a degenerate soliton tau function
of a (dual) TL hierarchy, where the variables $ \tilde t_k, \tilde{t'}_k$
play the role of higher times, see \cite{hypsol} for details.

Another example: if we take
${\bf t}={\bf t}(\infty,q)$ and choose
the collection $\xi_0$, $\xi_1$, $\xi_2,\dots$
by $\xi_m=\xi(\btt,q^m)-\xi({\btt'},q^{-m})$, then we obtain another soliton tau function of the
dual TL equation:
\begin{equation}
1+\sum_{k=1}^\infty\sum_{\scriptstyle\alpha_1>\cdots>\alpha_k\ge0 \atop
\scriptstyle n>\beta_1>\cdots>\beta_k\ge0}^\infty
\frac{\prod^k_{i<j}(q^{\alpha_i+1}-q^{\alpha_j+1})(q^{-\beta_j}-q^{-\beta_i})}
{\prod_{i,j=1}^k(q^{-\beta_i}-q^{\alpha_j+1})}\frac
{s_{(\alpha|\beta)}({\bf t'})}
{\prod_{i=1}^k(q;q)_{\alpha_i}\prod_{i=1}^k (q;q)_{\beta_i}}
e^{\sum_{i=1}^k
(\xi_{n+\alpha_i}-\xi_{n-\beta_i-1})}
\end{equation}

The partition functions of our discrete matrix models
(\ref{1MMdiscreteaxial-3}),(\ref{discreteMMBaxial}),
(\ref{discrete1MMC}), and (\ref{discrete1MMD}) are also soliton
tau functions.
This fact is clear after re-writing sums using the
Frobenius notations.

It will be interesting to compare these results with
\cite{SL}, where some models of random matrices where
interpreted as a limit of infinite-soliton tau function
in a different way.

\section{Two-matrix models with generalized interaction term}

\paragraph*{(A) Development of determinants in the Schur functions}
Consider the power series
\begin{equation}\label{f=sumexi-xizn}
f(z)=\sum_{n=0}^\infty e^{\xi_n-\xi_0}z^n
\end{equation}
\begin{Lemma} Let $x=(x_1,\dots,n),\ y=(y_1,\dots,y_n)$.
Then
\begin{equation}\label{Determinant-Wick-Fay}
c_n\frac{\det(f(x_iy_k))_{i,k=1,\dots,n}}{\Delta(x)\Delta(y)}=
\sum_{\lambda\in P;\ell(\lambda)\le n}
e^{\sum_{i=1}^n(\xi_{h_i}-\xi_{n-i})}
\underline s_\lambda(x)\underline s_\lambda(y)
\end{equation}
where $\Delta(x)=\prod_{1\le i<j\le n} (x_i-x_j)$,
$ 
c_1=1$, and $c_n =\prod_{i=0}^{n-1}e^{\xi_0-\xi_i}$ for $n>1$.
\end{Lemma}

Putting $a_n:=e^{\xi_n-\xi_0}$ so that
$e^{\sum_{i=1}^n(\xi_{h_i}-\xi_{n-i})}=c_n\prod_{i=1}^na_{h_i}$,
defining $n\times\infty$ matrix $M$ by $M_{ij}=a_jx_i^j$ and
$\infty\times n$ matrix $N$ by $N_{jk}=y_k^j$, and using
(\ref{detSc}), we observe that (\ref{Determinant-Wick-Fay}) reduces to
the following well-known formula for the determinant of product of
two matrices:
$$
\det(MN)=\sum_{h_1>h_2>\cdots>h_n\ge1}
\det(M_{1,\dots,n}^{h_n,\dots,h_1})
\det(N_{h_n,\dots,h_1}^{1,\dots,n})\,.
$$
While this may be the simplest proof of the lemma, the following
proof based on the known facts about tau functions is equally simple:
First, we identify $f(x_iy_j)$ with the tau function
\begin{equation}\label{tau(xy)}
\tau_r(1,x_i,y_j)=1+r(1)x_iy_j+
r(1)r(2)x_i^2y_j^2+\cdots
\end{equation}
by $r(k)=e^{\xi_k-\xi_{k-1}}$.
It may be shown by Wick's theorem (or the determinantal Fay identity
\cite[Theorem 4.2]{AvM}) that
\begin{equation}\label{det2'}
\tau_r(n,{\bf t}(x),{\bf t^*}(y))= c_n\frac{\det(
\tau_r(1,x_i,y_j))_{i,k=1}^n}{\Delta(x)\Delta(y)} \ ,
\end{equation}
where
\begin{equation}\label{cn1}
c_1=1\ ,\qquad c_n=\prod_{k=1}^{n-1} r(k)^{k-n}\ ,\quad n>1
\end{equation}
It means that for any power series $f$ as in (\ref{f=sumexi-xizn}) we have
$$
c_n\frac{\det(f(x_iy_k))_{i,k=1,\dots,n}}{\Delta(x)\Delta(y)}=
\tau _r(n,{\bf t}(x^{(n)}),{\bf t'}(y^{(n)})):=\sum_\lambda
e^{\xi_{h_i}-\xi_{n-i}}s_\lambda({\bf t}(x^{(n)}))
s_\lambda({\bf t'}(y^{(n)}))
$$
(Here $c_n^{-1}\Delta(x)\Delta(y)$ is
assumed to be non-vanishing; otherwise formula (\ref{det2'})
should be modified.)

\begin{remark} Choosing
$r(k)=(k+n-M)^{-1}\prod_{i=1}^p(k+a_i)\prod_{i=1}^s(k+b_i)^{-1}$
for the tau function of hypergeometric type, we obtain the {\em
hypergeometric functions of two matrix arguments} $X,Y$ \cite{V},
\cite{pd22}:
\begin{eqnarray}\label{taushurmathcalF2}
{}_p{\mathcal{F}}_s\left.\left(a_1+M,\dots,a_p+M\atop b_1+M,
\cdots,b_s+M\right| X,Y\right)=
\tau_r(M,{\bf t}(x^{(n)}),{\bf t}^*(y^{(n)}))=\nonumber\\
\sum_{\scriptstyle\lambda\atop\scriptstyle\ell(\lambda)\le n}
\frac{\prod_{k=1}^p
s_\lambda({\bf t}(a_k+M,1))} {\prod_{k=1}^s
s_\lambda({\bf t}(b_k+M,1))}
\left(s_\lambda({\bf t}_\infty)\right)^{s-p+1}\frac
{s_\lambda(X)s_\lambda(Y)}{s_\lambda(I_n)}
\end{eqnarray}
Here $x^{(n)}=(x_1,\dots,x_n)$ and $y^{(n)}=(y_1,\dots,y_n)$ are
the eigenvalues of the matrices $X$ and $Y$, respectively.
\end{remark}
Examples of ${}_p{\mathcal{F}}_s$\,: \ ${}_0{\mathcal{F}}_0(X,I_n)
=e^{\Tr X}$\,, ${}_1{\mathcal{F}}_0(a|X,I_n)=\det(I_n-X)^{-a}$\,.
Example of (\ref{det2'}):
\begin{equation}\label{detF2}
{}_p{\mathcal{F}}_s\left.\left(a_1+n,\dots ,a_p+n\atop b_1+n,
\cdots ,b_s+n\right| X,Y\right)=
\frac{c_n}{\Delta(x)\Delta(y)} \det\left({}_pF_s\left.\left(a_1+1,\dots
,a_p+1\atop b_1+1, \cdots ,b_s+1\right| x_iy_j\right)\right)_{i,j=1}^n\ ,
\end{equation}
which relates hypergeometric function of matrix arguments
to hypergeometric functions of one variable.

\paragraph*{(B) Angle integration of determinants}

We shall exploit the following formulae for integrations of Schur
functions over the unitary group \cite{Mac}. Let $d_*U$ be the
normalized Haar measure on $U(n)$, and let $\delta_{\mu,\lambda}$
be the Kronecker symbol. By $I_n$ we shall denote $n$ by $n$ unit
matrix. Let $A$ and $B$ be $n$ by $n$ matrices with the
eigenvalues $a_1,\dots,a_n$ and $b_1,\dots,b_n$ respectively. (We
are interested mainly in hermitian and unitary matrices, where
eigenvalues are, respectively, real numbers or numbers on the unit circle). Then
\begin{equation}\label{sAUBU^+}
\int_{U(n)}s_\lambda(AUBU^\dag)d_*U=
\frac{s_\lambda(A)s_\lambda(B)}
{s_\lambda(I_n)} \ ,
\end{equation}
and then consider the following model of two hermitian (or two
unitary) random matrices
\begin{equation}\label{multinew}
J_n=\int e^{\Tr \tilde V_1(M_1)+\Tr \tilde V_2(M_2)}g(M_1M_2)dM_1dM_2\,,
\end{equation}
where the measures $dM_1$ and $dM_2$ are defined in a standard way
(see \cite{Mehta}), and where the interaction term $g$ is a
hypergeometric tau function (see (\ref{exex'})) of the form
\begin{equation}\label{Gtau}
g(X)=\tau_r(n,{\bf t}(X) ,{\bf t}(I_n)):=\sum_{\lambda \in P \atop
\ell(\lambda)\le n} e^{\xi_{h_i}-\xi_{n-i}}s_\lambda(X)
s_\lambda(I_n) \ ,
\end{equation}
where $r(k)=e^{\xi_k-\xi_{k-1}}$ is not fixed, and where
\begin{equation}\label{t(X)}
{\bf t}(X) = (\frac11\Tr X,\frac12\Tr X^2,\frac13\Tr X^3,\dots)
\end{equation}

Then due to (\ref{sAUBU^+}) it is possible to perform the angle
integration over $U(n)$, where $M_1=U_1AU_1^\dag$,
$M_2=U_2BU_2^\dag$,
$A=\diag(a_1,\dots ,a_n)$, $B=\diag(b_1,\dots ,b_n)$.
\begin{equation}\label{Gangle}
\int_{U(n)}g(M_1M_2)d_*U=\tau_r(n,{\bf t}(M_1) ,{\bf t}(M_2))
\end{equation}

Finely we obtain the following integral over
eigenvalues $a_i$ and $b_i$:
\begin{eqnarray}
J_n&=&c_n\int \Delta(a)^2\Delta(b)^2\frac{\det f(a_ib_k) }{ \Delta(a)\Delta(b)}
\prod_{i=1}^n e^{\tilde V_1(a_i)+ \tilde V_2(b_i)} da_idb_i\nonumber\\
&=&
n!c_n\int \Delta(a)\Delta(b)
\prod_{i=1}^nf(a_ib_i)e^{\tilde V_1(a_i)+ \tilde V_2(b_i)} da_idb_i
\label{eigengenmulti}
\end{eqnarray}

\section{Conclusion}

Starting with the partition function of the model of normal
matrices and using the duality relations for Schur functions, we
get a Zoo of discrete models. In axial symmetric case these
discrete models may be identified with soliton tau functions. The
next step is to study the $n \to \infty$ limit of these models.
Our results means that we should obtain a behavior common for
groups of different models. Let us notice that discrete models
$Z_1-Z_5$, we have obtained, can be studied by the method of
orthogonal polynomials. It will be interesting to consider
continuous vs. discrete models in the framework of different
duality problems, see \cite{HIts}. These topis together with a
generalization of all results to multi-matrix models we shall
study in a more detailed paper.

\section{Acknowledgements}

The authors thank Oleg Zaboronsky and, most of all, John Harnad
for helpful discussions. A.O. thanks O. Zaboronsky, J. Harnad, and
J.J. Nimmo for the organization of his visits respectively to
Warwick, Oxford and Glasgow Universities in May--June 2003, where
this work was started. The work was supported by the Russian
Foundation for Fundamental Researches (Grant No 02-02-17382), the
Program of Russian Academy of Science ``Mathematical Methods in
Nonlinear Dynamics'', and a grant-in-aid for the scientific
research of Japanese Ministry of Education and Science.


\begin{thebibliography}{99}

\bibitem{Chau} Chau, L-L. and Yu, Y.: {\em Unitary polynomials in normal
matrix model and wave functions for the fractional quantum Hall effect},
Phys. Lett. {bf A167} p. 452 (1992)

\bibitem{Zabor1} Chau, L-L. and Zaboronsky, O.: {\em Normal matrix model,
Toda lattice hierarchy, and the two-dimensional electron gaz in the strong
magnetic field}, Proceedings in memory of professor Wolfgang Kroll, ed. J.P.Hsu
et. al., World Scientific, Singapore, 1997

\bibitem{Zabor} Chau, L-L. and Zaboronsky, O.: On the Structure of
Correlation Functions in the Normal Matrix Models, Commun. Math. Phys.
{\bf 196} (1998) 203--247; {\em hep-th/9711091}

\bibitem{MWZ} Mineev-Weinstein, M., Wiegmann, P. and Zabrodin, A.:
Integrable structure of interface dynamics, {\em Phys. Rev.
Lett.} {\bf 84} (2000) 5106--5109; {\em nlin.SI/0001007}

\bibitem{Tinit} Takasaki, K.: Initial value problem for the
Toda lattice hierarchy, {\em Adv. Stud. Pure Math.} {\bf 4}
(1984) 139--163

\bibitem{TI} Takebe, T.: Representation Theoretical Meaning
of Initial Value Problem for the Toda Lattice Hierarchy I, {\em
LMP\/} {\bf 21} (1991) 77--84

\bibitem{TII} Takebe, T.: Representation Theoretical Meaning
of Initial Value Problem for the Toda Lattice Hierarchy II, {\em
Publ. RIMS, Kyoto Univ.} {\bf 27} (1991), 491--503

\bibitem{TT} Takebe, T. and Takasaki, K.: Integrable Hierarchies
and Dispersionless Limit, {\em Rev. Math. Phys.} {\bf 7} (1995)
743--808; {\em hep-th/94050096}

\bibitem{Mac} Macdonald, I.G.: {\em Symmetric Functions and Hall
Polynomials}, Second edition, Clarendon Press, Oxford, 1995

\bibitem{ZSh} Zakharov, V.E. and Shabat, A.B.: {\em J. Funct.
Anal. Appl.} {\bf 8} (1974) 226

\bibitem{DJKM} Date, E., Jimbo, M., Kashiwara, M. and Miwa, T.:
Transformation groups for soliton equations. In:
Jimbo, M. and Miwa, T. (eds) {\em Nonlinear integrable systems---%
classical theory and quantum theory\/} pp. 39--120, World
Scientific, 1983

\bibitem{JM} Jimbo, M. and Miwa, T.: Solitons and Infinite
Dimensional Lie Algebras, {\em Publ. RIMS Kyoto Univ.}
{\bf 19} (1983) 943--1001

\bibitem{D} L. A. Dickey, {\em Soliton Equations and
Hamiltonian Systems}, World Scientific, Singapore, 1991

\bibitem{AM} Mikhailov, A.V.: On the Integrability of two-dimensional
Generalization of the Toda Lattice, {\em Letters in Journal of
Experimental and Theoretical Physics} {\bf 30} (1979) 443--448

\bibitem{UT} Ueno, K. and Takasaki, K.: {\em Adv. Stud. Pure
Math.} {\bf 4} (1984) 1--95

\bibitem{GMMMO} Gerasimov, A., Marshakov, A., Mironov, A.,
Morozov, A., and Orlov, A.Yu.: Matrix Models of 2D Gravity and Toda
Theory, {\em Nuclear Physics B\/} {\bf 357} (1991) 565--618

\bibitem{ZKMMO} Zabrodin, A., Kharchev, S., Mironov, A., Marshakov, A.
and Orlov, A.: Matrix Models among Integrable Theories: Forced
Hierarchies and Operator Formalism, {\em Nuclear Physics B\/}
{\bf 366} (1991) 569--601

\bibitem{pd22} A.Yu. Orlov and D.M. Scherbin ``Multivariate
hypergeometric functions as tau functions of Toda lattice and
Kadomtsev-Petviashvili equation'', Physica D {\bf 152--153} (2001)
51--56

\bibitem{hypsol} Orlov, A.Yu.: Rational solutions of KP hierarchy
as multisoliton solutions of a dual KP (and TL) hierarchy
{\em submitted to Phys. Lett. A};
Hypergeometric tau
functions $\tau({\bf t},T,{\bf t}^*)$ as $\infty$-soliton tau
function in $T$ variables; {\em solv-int/0305001}

\bibitem{1'} Orlov, A.Yu.: New Solvable Matrix Integrals,
 International Journal of Modern Physics A Vol. 19, Supplement
 (2004) 276-293; {\em nlin. SI/0209063}

\bibitem{galipolli} A. Yu. Orlov, ``New Solvable Matrix Models---$U(n)$ case''
Proceedings of the workshop
Nonlinear Physics: Theory and Experiment. II,
Universit\`a di Lecce--Consortium Einstein, 27 June--6 July 2002,
Gallipoli, Italy; editors: M.J. Ablowitz,
M. Boiti, F. Pempinelli, B. Prinari; World Scientific New Jersey,
London, Singapore, Hong Kong (2002) pp. 99--100

\bibitem{CRM} Harnad, J. and Orlov, A.Yu.:
Schur Function Expansions of Matrix Integrals, preprint CRM 2001

\bibitem{Cadiz} Harnad, J. and Orlov, A.Yu.: Scalar products of symmetric
functions and matrix integrals, {\em nlin.SI/0211051}. (To appear
in proceedings of NEEDS2002, eds., A. Gonzales, World Scientific,
2002.) Theoretical and Mathematical Physics {\bf 137} No 3 (2003)
1676--1690

\bibitem{Kos1} Kostov, I.K., Staudacher, M. and Wynter, T.:
Complex Matrix Models and Statistics of Branched Covering of 2D
Surfaces, {\em Commun. Math. Phys.} {\bf 191} (1998) 283--298; {\em
hep-th/9703189}

\bibitem{Kos2} Kostov, I.K.: Exact Solution of the Six-Vertex
Model on a Random Lattice, {\em Nucl. Phys. B} {\bf 575} (2000)
513--534; {\em hep-th/9911023}

\bibitem{ZM} Zakharov, V.E., Manakov, S.V., Novikov, S.P. and
Pitaevsky, L.P. {\em The Theory of Solitons.
The Inverse Scattering Method}, Moscow, Nauka, 1980

\bibitem{V} Vilenkin, N.Ya. and Klimyk, A.U.:
{\em Representation of Lie Groups and Special Functions.
Volume 3:
Classical and Quantum Groups and Special Functions}, Kluwer
Academic Publishers, 1992

\bibitem{SL} Spiridonov, V. and Loutsenko, I.M.: Soliton Solutions
of Integrable Hierarchies and Coulomb Plasmas,
{\em Journal
of statistical physics} {\bf 99} (2000) 751--767; {\em
cond-mat/9909308}

\bibitem{AvM} Adler, M. and van Moerbeke, P.: The spectrum of
coupled random matrices, Annals of Math., {\bf 149}, 921--976 (1999)

\bibitem{BB} Borodin, A. and Boyarchenko, D.: Distribution
of the first particle in discrete orthogonal polynomial ensembles
{\em math-ph/0204001}

\bibitem{KMM} S. Kharchev, A. Marshakov and A. Mironov and A. Morozov,
Int. J. Mod. Phys. {\bf A10} 2015 (1995), ``Generalized
Kazakov-Migdal-Kontsevich Model: group theory
aspects'', hep-th/9312210 ; Nucl. Phys. {\bf B397} 339 (1993);
hep-th/9203043

\bibitem{Olshanski} Borodin, A. and Olshanski, G.: Random
partitions and Gamma kernel, {\em math-ph/0305043}

\bibitem{Mehta} Mehta, M.~L., {\it Random Matrices}, 2nd edition (Academic, San
Diego, 1991).

\bibitem{HIts} Harnad, J. and Its, A.: Integrable Fredholm Operators
and Dual Isomonodromic Deformations'', Commun. Math. Phys. 226,
497--530 (2002); solv-int/9706002

\end{thebibliography}
\end{document}